# Comments on "Negative Specific Heat in Astronomy, Physics and Chemistry"

H. Umirzakov
Institute of Thermophysics, Russia, 630090, Novosibirsk, Lavrentev prospect, 1.
e-mail: cluster125@gmail.com

## Abstract

It is shown that the "proof" of [1] that the specific heat of the system of classical point particles interacting with each other via uniform gravitational potential energy may be negative is incorrect.

The "proof" that the isochoric heat capacity (specific heat) $C_V$ of the system of particles interacting with each other via uniform gravitational potential energy may be negative was published in [1]. The "proof" consists of following:

*"The Virial theorem for a steady state under a potential energy, $U$, that scales like $r^{-n}$ reads [for external (or edge) pressure $p_e$ and volume $V = 4\pi r_e^3/3$],*

$$2K + n \cdot U = 3 p_e \cdot V. \qquad (1)$$

*For gravity $n=1$ and the total energy, $E$, is the sum of kinetic and potential parts. So for an isolated gravitational system ( $p_e = 0$ )*

$$E = -K < 0 \qquad (2)$$

*but for particles in motion $K = 3NkT/2$ so*

$$dE/dT = C_V = -3Nk/2, \qquad (3)$$

*which is clearly negative!"*

As one can see the "proof" implicitly implies that the system consisting of $N$ classical point particles is considered and "the virial theorem" is valid for the system. According to "the virial theorem" [2-5] the velocities of particles must be not equal to infinity. But the velocities of gravitating point particles can be equal to infinity [5]. Therefore "the virial theorem" is not valid.

The equation (1) is incorrect because according to "the virial theorem" [2-4] the potential energy of interaction between particles $U$ in Eq. (1) must be replaced by its time average.

The "proof" is incorrect, because equation (2) is valid along the isobar $p_e = 0 = const$, so the equation (3) must be replaced by $(\partial E/\partial T)_p = C_P = -3Nk/2$, where $C_P$ is the isobaric heat capacity.